\documentclass {article}
\usepackage[dvips]{graphicx}
\bigskip \bigskip
\begin{document}

\bigskip \bigskip
\centerline{\bf ENERGY OF INTERACTION BETWEEN SOLID SURFACES AND
LIQUIDS }\medskip \vskip 0,5 cm \centerline{\bf Henri
GOUIN}\medskip
\bigskip
\centerline{ L. M. M. T. \,    Box 322,\,  University of
Aix-Marseille}\medskip \centerline{ Avenue Escadrille
Normandie-Niemen, 13397 Marseille Cedex 20 France}\medskip
\bigskip

\centerline{\small E-mail:  henri.gouin@univ-cezanne.fr}
\bigskip

\bigskip \bigskip
\centerline{\bf Abstract}\bigskip

We consider the wetting transition on  a planar surface in contact
with a semi-infinite fluid. In the classical approach, the surface
is assumed to be solid,
 and when interaction
between solid and fluid is sufficiently short-range, the
contribution of the fluid can be represented by a surface free
energy with a density of the form \ $\Phi (\rho_S)$, where $\rho_S$
\ is the limiting density of the fluid at the surface.\par

In the present paper we propose  a more precise representation of
the surface energy that takes into account not only the
 value of  $\rho_S$   but also the contribution from the whole density
 profile
$\rho(z)$ of the
 fluid, where   $z$  is coordinate normal to the surface.

\noindent The specific value of the functional of  $\rho_S$  at the
surface is expressed in mean-field approximation through the
potentials of intermolecular interaction and some  other parameters
of the fluid  and the solid wall.

\noindent An extension to the case of fluid mixtures in contact with
a solid surface is proposed.

\bigskip \bigskip  \bigskip \bigskip

\medskip {\bf 1. Introduction }
\medskip

The phenomenon  of surface wetting is a subject of many experiments.
They have already been used  to determine
 many important properties of the wetting behavior for liquids
on low-energy solid surfaces $^1$. To gain a theoretical
explanation of the
 phenomenon of wetting, a
generalized van der
 Waals model is often used $^2$.

\noindent In the 1950's Zisman $^3$ developed an experimental method
of characterizing the free energy of a solid
 surface by the measurement of
the contact angle with respect to the liquid-vapor surface tension
and by changing the test liquid. More recent measurements done by Li
and all $^{4,5}$ improved the understanding  of
 this problem.
While the contact angle and surface tension are macroscopic
quantities, they have their origin in molecular interactions.

\noindent In the following, we use a mean-field theory to
investigate how  the surface energy is related to
 molecular interactions. The approximation of mean field theory  is too simple to
be quantitatively accurate. However it does
 provide a qualitative understanding and allows one to calculate explicitely the
magnitude of the coefficients  in our model.

\noindent In 1977, John Cahn$^1$ gave simple illuminating arguments
to describe the interaction between solids and liquids. His model is
based on a generalized van der Waals theory of fluids treated as
attracting hard spheres$^6$. It entailed assigning to the solid
surface an energy that
 was a functional of the liquid density "at the surface"; the particular form
  $ \displaystyle {\Phi (\rho_S) \ = \ - \gamma_1 \ \rho_S +
{\gamma_2\over 2} \ \rho^2_S }$\ of this energy,  where $ \rho_S $
is the fluid density at the solid wall, is now widely known in the
literature and is due to Nakanishi and Fisher $^7$. It was
thoroughly examined in a review paper by de Gennes $^8$. To account
for the wetting behavior of liquids on solid surfaces, one needs to
know \ $\Phi (\rho_S)$; a major object of the present paper is to
obtain explicit formulas for the coefficients $ \gamma_1$   and $
\gamma_2 $, expressing them in terms
 of the parameters in assumed microscopic interaction potentials.

\noindent Three hypotheses are implicit in Cahn's picture.

$i)$ \ For the liquid density to be taken to be a smooth function
$\rho(z)$    of the distance   $z$   from the solid surface, that
surface is assumed to be flat  on the scale of molecular sizes and
the correlation length is assumed to be greater
 than intermolecular distances (as is the case, for example, when the
 temperature
$T$  is not far from the critical temperature $T_c$).

$ii)$\ The forces between solid and liquid are of short range
 with respect to
 intermolecular distances.

$iii)$ \ The fluid  is  considered in the framework of a mean-field
theory. This
 means, in particular, that the free energy of the fluid is a classical
so-called "gradient square functional".

\noindent The point of view that the fluid in the interfacial region
may be treated as  bulk phase with a local free-energy
 density and an additional contribution arising from the
nonuniformity which may be approximated by a gradient
 expansion truncated at the
second order is most likely to be successful and perhaps even
quantitatively accurate near the critical point $^6$. \noindent Some
numerical calculations based on Cahn's model and comparison with
experiments can be found
 in the literature $^{2,9}$.  \par
\noindent The aim of this paper, as was stated, is  to obtain the
values of the different coefficients of the expression of energy
due to the interaction between a solid wall and a liquid bulk. These
values are
 associated with
intermolecular potentials of the liquid and the solid wall. They
take
 into account
the molecular sizes and finite range of interactions between the
molecules of the liquid and the solid. The value of coefficients
$\gamma_1$   and   $\gamma_2$   are positive. Consequently the
$\gamma_1$   term
 describes an attraction of the liquid by the solid, and
 the   $\gamma_2$   term
a  reduction of the  attractive interactions near the surface. In
fact, the calculation leads to a  more complex functional dependence
than the one given by Cahn, Nakanishi and Fisher or de Gennes
$^{1,7,8}$.
 The energy of the wall
 takes into account not only
the value of the density at the wall but also the gradient of
density normal
 to
the wall. This should be more accurate when the variation of density
is strong  with respect to
 molecular sizes. The density of the liquid   $\nu_1$  is  chosen depending on three coordinates
$ x, y, z$. The direction to the wall is associated with $z$, and
near the wall, the density may be chosen to be a function only of
$z$. In fact such an assumption does not simplify the calculations.
In our expression, the energy  smoothly depends  on $x, y$ and
allows continuous variations of density along the wall. The
connection with Nakanishi-Fisher expression is not affected by this
more general hypothesis.  We denote by $E_S$ the new expression of
the wall energy.\medskip

\noindent The calculation may be extended to more complex cases: for
example to the case of a liquid mixture in
 contact with a solid wall.  We propose a general expression of
the energy density at the solid
 surface. We note that the distribution of
the concentrations of the components may be influenced
 by a solid wall
effect. \bigskip
\bigskip

{\bf 2. Description of the solid-fluid  interaction  in mean field
approximation}
\medskip

 In  regions of a fluid where the density
$\rho $   in nonuniform, the intermolecular potentials produce a
force
 on a given molecule that may generate surface
tension effects $^{6,10}$. Using classical molecular theory, it is
possible to obtain a system of pressure and capillary tension that
is mathematically equivalent to
 the stress tensor of a continuum model $^{6,11}$.
 Such a description does not take into account the possibility
 of interaction of the liquid with the solid wall (when the
distribution of density varies
 strongly in the direction normal to
a solid surface).\par \noindent   These effects constitute the
subject of the present paper.\medskip

\noindent  The effects are illustrated in Figure 1 and are described
below. We consider a flat  solid wall. The approach is aimed to
describe  the interaction of a macroscopic  surface
 with a  bulk fluid.
The so-called "hard sphere diameter" of the molecules is denoted by
\ $\sigma$ \ for the fluid and \ $ \tau $ \ for the solid.
 Then, the minimal distance between solid and
liquid molecules is \ $\displaystyle \delta \ = \ {1\over 2}\
(\sigma+\tau )$.

\noindent In mean-field theory, we represent by \ $\psi (r)$ \ the
intermolecular potential between
 two molecules of the fluid and respectively \ $\chi(r)$ \  the potential between a fluid molecule and a wall
molecule at separation \ $r$. The density of molecules at a point
depends on its coordinates \ $\ x  ,  \ y, \ z $, but the masses \
$m_1$ \ and \ $m_2$ \  of each molecule of the liquid and solid are
assumed to be fixed. The energy corresponding to the action of all
molecules of the liquid and the solid
 on a given molecule 1 located
in \, $0$\ (see Figure 1) is
$$W_0 \ =  \ \sum_i \ m^2_1 \ \psi (r_i) + \sum_j \ m_1 m_2 \ \chi(r_j). \eqno (1)$$
\begin{figure}[h]
\begin{center}
\includegraphics[width=5cm]{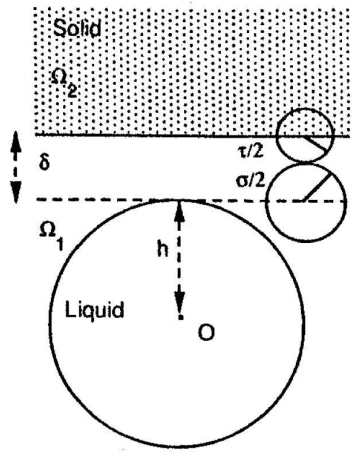}
\end{center}
\caption{\emph{Molecular layer between the liquid and the solid surface.  }}\label{fig1}
\end{figure}

The first summation (over i) is over fluid molecules (except for
molecule $1$), and the second summation (over j) is over the wall
molecules. Molecule 1 is in the fluid and \ $r_i ,  r_j$
 are  distances from molecule \ $i$ \ or  \ $j$ \ to
molecule $1$.  \medskip

\noindent Denoting by \ $\nu_1 d \omega_1$ \ the number of molecules
of fluid in the volume element \ $d\omega_1$ \ and \ $\nu_2 d
\omega_2$\ the number
 of molecules of solid in
 volume element \ $d\omega_2$, the
potential energy resulting from the action of all  molecules in the
medium on molecule
  $1$ located in \, $0$ \, may be expressed in a continuous way:
$$W_0\ = \ \int\!\!\!\!\int\!\!\!\!\int_{\Omega_1} \ m^2_1 \ \psi(r) \nu_1 \
d\omega_1 \ + \ \int\!\!\!\!\int\!\!\!\!\int_{\Omega_2} \ m_1 m_2 \
\chi(r) \nu_2 \ d\omega_2   \eqno (2)$$\par \noindent where \
$\Omega_1 $ \ and \ $\Omega_2 $ \ are the domains occupied by the
liquid and the solid.\par

\noindent Potential energy \ $W_0$ \ will be summed over all the
molecules of the liquid. So, the first integral  is counted twice
 in the previous integral over the
domain  \ $\Omega_1 $ \ occupied by the liquid. In such a condition,
we
 have to
consider only the expression  \par

$${1\over 2} \ \int\!\!\!\!\int\!\!\!\!\int_{\Omega_1} \ m^2_1 \ \psi(r) \nu_1 \
d\omega_1 \ + \ \int\!\!\!\!\int\!\!\!\!\int_{\Omega_2} \ m_1 m_2 \
\chi(r) \nu_2 \ d\omega_2 \eqno (3)$$\par \noindent for the
potential energy \ $W_0$.\par \noindent
 Let \ $\nu_1(x , y ,  z )$  be an analytic
function in each point  \ $M (x , y ,  z )$  of the liquid.
\par
\noindent In the next derivation we replace at the point $(0,0,0) \
\nu_1 $ with its  Taylor expansion in variables $ x,\ y,\ z$ \
limited to the second order. Such an assumption means that the
forces between the solid and the liquid are of a short range. This
is similar to the expansion given by Rocard$^{11}$. This means that
$\psi(r)$ is a rapidly decreasing function of $r$. Then, it is only
necessary to know the distribution of molecules at a short distance
from molecule 1. This case reflects the  reality when the main force
potentials decrease as $r^{-6}$ with the distance. So, the
potentials decrease very rapidly from the solid wall. The
distribution of density inside the solid is  assumed to be uniform.
We write this expansion:
$$ \nu_1 (x , y ,  z ) \ = \ \nu_{10} + x  \ {\partial \nu_{10}\over \partial x} \
+ y \  {\partial \nu_{10}\over \partial y} \ + z \  {\partial
\nu_{10}\over \partial z} \ + \eqno (4)$$
$$ {1\over 2} \ \bigg(\ x^2 \
{\partial^2 \nu_{10}\over \partial x^2 } \ + y^2  \  {\partial^2
\nu_{10}\over \partial y^2 } \ + z^2  \  {\partial^2  \nu_{10}\over
\partial z^2 } \ + \ 2 x y \ {\partial^2  \nu_{10}\over \partial x
\partial y } \ + \ 2 x z {\partial^2  \nu_{10}\over \partial x
\partial z } \ + \ 2 y z {\partial^2  \nu_{10}\over \partial y
\partial z } \bigg) \ + . . .$$ \noindent where \ $ \displaystyle
\nu_{10} \ , \  {\partial \nu_{10}\over \partial x} \ ,
 {\partial^2 \nu_{10}\over \partial x^2} \ , \ldots $ \ represent the values of \
$\nu_1$ \ and its derivatives at point \ $ (0,0,0) $ .\par \noindent
If we note \ $\rho_1 \ = \ m_1 \nu_{10}$ \ and
 $\rho_2 \ = \ m_2 \nu_2$ \ the densities in the
liquid at point \ $(0,0,0)$ \ and inside the solid wall, we obtain
\par

$$\displaystyle W_0 \ = \ 2\pi\  m_1
 \rho_1 \int^{\infty}_{\sigma}\ r^2 \psi(r) dr + \pi \ m_1 \rho_1\int^{\infty}_h\
r (h-r) \psi (r) dr \ + $$
$$\displaystyle 2\pi \ m_1 \rho_2 \int^{\infty}_h \ r (r-h) \chi (r) dr +
 \ {\pi\over 2} \ m_1
 \ {\partial \rho_1\over \partial z}\ \int^{\infty}_h\
r (h^2 - r^2) \psi (r) dr \ + $$
$$\displaystyle {\pi\over 3} \ m_1 \Delta \rho_1  \int^{\infty}_{\sigma}\ r^4
\psi(r) dr + \ {\pi\over 12} \ m_1 \Delta_T \rho_1 \int^{\infty}_h \
(-2 r^4 + 3 h r^3 - r h^3) \psi (r) dr  \ + $$
$$\displaystyle  {\pi\over 6} \ m_1 \ {\partial^2 \rho_1\over
\partial z^2} \int^{\infty}_h \ r (h^3 - r^3) \psi (r) dr \
\eqno (5)$$

\noindent
 In Eq. (5), \ $h$ \ represents the distance between the molecule $1$ and the solid wall, \ $\Delta$ \
is the
 Laplace operator and
$\Delta_T$
 \ is the Laplace operator tangential to the wall.\par
\noindent Details of the calculation are given in Appendix 1.\par
\noindent Notice that the last two terms may also be written in the
form
$$\displaystyle {\pi\over 12} \ m_1 \Delta \rho_1 \int^{\infty}_h \
(-2 r^4 + 3 h r^3 - r h^3) \psi (r) dr + {\pi\over 4} \ m_1
{\partial^2 \rho_1\over \partial z^2}\ \int^{\infty}_h\ rh (h^2 - r
^2) \psi (r) dr \eqno (6)$$ \noindent The energy density per unit
volume at point \ $0 $ \ is \ $E_o \ = \ \nu_{10} \ W_0$. \par
\noindent In fact, this result is independant of the reference
point. If we denote \ $E$ \ the energy per unit volume at any point
\ $M$ \ in the liquid, we obtain  \par

$$\displaystyle E \ = \ 2\pi\ \rho^2_1
 \int^{\infty}_{\sigma}\ r^2 \psi(r) dr + \pi \ \rho^2_1 \int^{\infty}_h\
r (h-r) \psi (r) dr \ + $$
 $$\displaystyle 2\pi \ \rho_1 \rho_2 \int^{\infty}_h \ r (r-h) \chi (r) dr +
 \ {\pi\over 2} \ \rho_1
 \ {\partial \rho_1\over \partial z}\ \int^{\infty}_h\
r (h^2 - r^2) \psi (r) dr \ + $$
$$\displaystyle {\pi\over 3} \ \rho_1 \Delta \rho_1  \int^{\infty}_{\sigma}\ r^4
\psi(r) dr + \ {\pi\over 12} \ \rho_1 \Delta_T \rho_1
\int^{\infty}_h \ (-2 r^4 + 3 h r^3 - r h^3) \psi (r) dr \ + $$
$$\displaystyle  {\pi\over 6} \ \rho_1 \ {\partial^2 \rho_1\over
\partial z^2} \int^{\infty}_h \ r (h^3 - r^3) \psi (r) dr  \ \eqno(7)
$$
\noindent Expression (7) yields
$$
E \ = \ K \rho^2_1 + \alpha (h) \rho_1 \rho_2 \ + \ \beta (h)
\rho^2_1 \ + \ \gamma (h) \rho_1\ {\partial \rho_1\over \partial h}
\ + \ K b^2 \rho_1 \Delta \rho_1 \ + \ B \eqno (8) $$
with
$$\displaystyle K\ = \ 2\pi \int^{\infty}_{\sigma}\
r^2 \psi(r) dr \ ,  \hskip 1 cm K b^2 \ = \ {\pi\over 3 }
\int^{\infty}_{\sigma}\ r^4 \psi(r) dr $$
$$\displaystyle \alpha (h) \ = \ 2 \pi
\int^{\infty}_h \ r (r -  h ) \chi (r) dr \ , \ \ \beta (h) \ = \
\pi \int^{\infty}_h \ r (h -  r ) \psi (r) dr $$
$$\displaystyle \gamma (h) \ = \ -\ {\pi\over 2} \
\int^{\infty}_h \ r (h^2 -  r^2 ) \psi (r) dr $$
$$\displaystyle B   =    {\pi\over 12} \, \rho_1 \Delta_T \rho_1
\int^{\infty}_h  (- 2 r^4 + 3 h r^3 - r h^3)\psi (r) dr    +
{\pi\over 6} \, \rho_1 \, {\partial^2 \rho_1\over
\partial z^2} \int^{\infty}_h   r (h^3 - r^3) \psi (r) dr  $$

\noindent The constant \ $b$ \ appearing in the $\rho_1 \Delta
\rho_1$ term denotes the covolume of the liquid as in the van der
Waals equation $^{11,12}$ .\par \noindent The corresponding energy
of the fluid is   {$$\displaystyle W \ = \
\int\!\!\!\!\int\!\!\!\!\int_{\Omega_1} \ E \ d \omega _1    \eqno
(9)$$} \noindent where\  $E $ \ is given by Eq. (8) .\par
\noindent To take  into account the kinetic effects like in Rocard
$^{11}$ , p. 392, we must add the terms corresponding to
 kinetic
pressure
 to the value of \ $W$. The first term \ $K\rho^2_1$ \ yields the internal
pressure.\par \noindent Now, we calculate the different values of
coefficients in Eq. 8 in a special case of London forces.
\bigskip

{\bf 3. Calculations of the energy of interaction in the case of
London forces}

\medskip
It is now possible to calculate the value of \ $W$ \ for  the
particular interaction potentials. For example, one can take (see
Appendix 2)
$$ \psi (r) \ = \ - {k\over r^{n-1}} \hskip 1,5 cm \hbox{and} \hskip 1,5 cm
\chi (r) \ = \ -{\mu\over r^{n-1}} \  \eqno (10)$$ \noindent In the
case of London forces one has $^{10} $   \ n=7. \par \noindent In
fact this rough approximation is valid only at short range from the
wall (hypothesis \ $ii$). Following the calculations in  Appendix 2,
we obtain that the two integrals  in the
 \ $B$ term are of the order of \ $\displaystyle {1\over h} $ ;
 moreover,
$$ \alpha (h) \ = \ -
{\mu \pi\over 6 h^3} \ , \ \beta (h) \ = \ { k \pi\over 12 h^3} \ ,
\ \gamma (h) \ = \ - {k \pi\over 8 h^2}\   \eqno (11)$$\par
\noindent Integrals associated with Eq. (9) are taken over an
interval \ $\lbrack \delta , \ d \rbrack $ , where \ $d$ \ is the
range of
 molecular forces in the liquid and $ \delta$ is the minimal distance between solid and
liquid molecules.\par \noindent For the same reasons as in Eq. (4),
the expansion of \
 $\displaystyle \rho_1\ , \ {\partial \rho_1\over \partial h} \ , \
{\partial^2 \rho_1\over \partial h^2} $ \ at the solid wall is taken
 in the
form
$$\displaystyle \cases {
\displaystyle \ \ \ \rho_1 \ = \ \rho_S + h \
 {\partial \rho_S\over \partial h}\ +  \
{h^2\over  2} \ {\partial^2 \rho_S\over \partial h^2}\ + \ \cdots
\cr\cr \displaystyle \ \ {\partial \rho_1\over \partial h}  \ = \
{\partial \rho_S\over
\partial h}\ +   \ h \ {\partial^2 \rho_S\over \partial h^2}\ + \ \cdots
 \cr\cr
\displaystyle \ \ {\partial^2 \rho_1\over \partial h^2}  \ = \
{\partial^2 \rho_S\over \partial h^2} \ +  \ \cdots }\eqno(12)$$

\noindent Terms \ $\displaystyle \rho_S \ , \ {\partial \rho_S\over
\partial h}\ , {\partial^2 \rho_S\over \partial h^2}\ $ denote
values of \ $\rho_1$ \  and its normal derivatives at the solid
wall.\par \noindent
 This means   $\displaystyle \rho_S \ = (\rho_1)_S , \ {\partial
\rho_S\over
\partial h} = ({\partial \rho_1\over \partial h} )_S ,\ {\partial^2 \rho_S\over \partial
h^2}\ = ({\partial^2 \rho_1\over \partial h^2})_S$. \par \noindent
Now we make the important assumption that the variations of $\rho_1$
take into account several molecular ranges. Hence, $\displaystyle{
\delta \over d}$ can be considered
 as a small parameter.
It implies that the first derivative of $\rho_1$ with respect to $h$
is on the order of $\displaystyle {\rho_1\over d} $, the second
derivative of $\rho_1$ is on the order of
 $\displaystyle {\rho_1\over d^2} $.
Then, the $B$ term can be removed from the integration of $E$.
Consequently, in the calculation of the surface energy in Eq. (4),
the second derivative terms
 may be removed   (but not in the calculation of the bulk energy of the liquid
associated with $ K (\rho^2_1 + b^2 \rho_1 \Delta \rho_1 )$). This
result is in agreement with some considerations by de Gennes
$^{13}$.

\noindent Keeping terms up to the first  order in Eq. (8) yields
$$\displaystyle E  =   K (\rho^2_1 + b^2 \rho_1 \Delta \rho_1 )+ \alpha (h) \rho_2
\bigg( \rho_S + h   {\partial \rho_S\over \partial h} \bigg) +
\beta (h)   \bigg(   \rho^2_S + 2 h \rho_S   {\partial \rho_S\over
\partial h}  \bigg) + \gamma (h)   \rho_S   {\partial \rho_S\over
\partial h}
$$
\noindent or
$$\displaystyle E  =  K (\rho^2_1 + b^2 \rho_1 \Delta \rho_1 )+ \alpha (h) \rho_2
\rho_S + \beta (h) \rho^2_S +h \alpha (h) \rho_2 \ {\partial \rho_S\over \partial h}\ +
\ {2 h \beta (h) + \gamma (h)\over 2} \ {\partial \rho^2_S\over
\partial h} \   $$

\noindent Then, the energy of the fluid is  \medskip

\centerline{$ W = W_1+W_2$ , \hskip 0,5 cm  with}\par $\displaystyle
W_1 \ = \ \int\!\!\!\!\int\!\!\!\!\int_{\Omega_1} \ K (\rho^2_1 +
b^2 \rho_1 \Delta \rho_1 ) d \omega _1 \ \   $

$\displaystyle W_2 \ = \ \rho_2 \rho_S \
\int\!\!\!\!\int\!\!\!\!\int_{\Omega_1} \ \alpha(h) dh dx dy \ + \
\rho^2_S\  \int\!\!\!\!\int\!\!\!\!\int_{\Omega_1} \ \beta (h) dh dx
dy \  $

\hskip 0,5 cm \ \ $ \displaystyle  + \ \rho_2 \ {\partial
\rho_S\over \partial h} \
 \int\!\!\!\!\int\!\!\!\!\int_{\Omega_1} \ h \alpha (h) dh dx dy+ \ {\partial \rho^2_S\over
\partial h} \ \int\!\!\!\!\int\!\!\!\!\int_{\Omega_1} \ \ {2 h \beta
(h) + \gamma (h)\over 2} \ dh dx dy  \ \  $

\noindent with $$\displaystyle \int^{+\infty}_{\delta} \ \alpha(h)
dh \ = \ - {\mu\pi\over 12 \delta^2}\ , \hskip 0,5 cm
\int^{+\infty}_{\delta} \ \beta(h) dh \ = \ {k \pi\over 24
\delta^2}$$

$$\displaystyle \int^{+\infty}_{\delta} \ h \alpha(h) dh \ = \ -
{\mu\pi\over 6 \delta}\  , \hskip 0,5 cm \int^{+\infty}_{\delta} \ h
\beta(h) dh \ = \ {k \pi\over 12 \delta}$$

$$\displaystyle \int^{+\infty}_{\delta} \ {\gamma (h)\over 2} \  dh \
= \ - {k \pi\over 6 \delta} \ \  \eqno(13) $$

\noindent We note that \ \ $\displaystyle \rho_1\Delta \rho_1 \ = \
\rho_1 \ div ( \overrightarrow{grad}\ \rho_1) \ = \ div  (  \rho_1 \
\overrightarrow{grad}\ \rho_1) - (\overrightarrow{grad}\
\rho_1)^2$\par \noindent and the Stokes formula yields
$$\displaystyle W_1 \ = \  \int\!\!\!\!\int\!\!\!\!\int_{\Omega_1} \
 K  ( \rho^2_1 - b ^2 (\overrightarrow{grad}\  \rho_1)^2 ) d\omega_1 \ + \
\int\!\!\!\!\int_{\Sigma} \ Kb^2 \rho_S \ {\partial \rho_S\over
\partial h} \ ds \eqno(14) $$ Here \ $\Sigma \ $ notes the surface of
the solid wall corresponding to the boundary of \ $\Omega_1$. As it
was said in paragraph 2, to take into account the kinetic effects,
we must
 add to energy \ $W_1$  the terms corresponding to kinetic pressure. \par
\noindent The first term \ $K \rho^2_1$ \ yields the internal
pressure, and consequently
$$\displaystyle \varepsilon \ = \ \alpha (\rho_1 , \eta ) \ - \ {K
b^2\over \rho_1} \ (\overrightarrow{grad}\  \rho_1)^2 $$ \noindent
denotes the internal specific energy. Term \ $\rho_1 \alpha (\rho_1
, \eta)$ \ is the bulk internal volume energy as a function of \
$\rho_1$ \ and of the specific entropy \ $\eta $ \ in the liquid.

\noindent Let us note that
$$ \int\!\!\!\!\int_{\Sigma}\ Kb^2  \rho_S \ {\partial\rho_S\over \partial h}\
ds \ = \ \int\!\!\!\!\int_{\Sigma}\ - {k\pi\over 6\sigma}\
{\partial\rho^2_S\over \partial h} \ ds$$\par \noindent Then, a
straighforward calculation yields the energy of the liquid in the
well known form
$$ W\ = \int\!\!\!\!\int\!\!\!\!\int_{\Omega_1}\ \Biggl\{ \ \rho_1 \alpha (\rho_1 ,
\eta ) - Kb^2 \ (\overrightarrow{grad}\, \rho_1)^2\ \Biggl\} d \omega_1 +
 \int\!\!\!\!\int_{\Sigma}\ E_S \ d s $$\par
\noindent but with
$$\displaystyle E_S \ = \ - \gamma_1 \ \rho_S + {\gamma_2\over 2} \ \rho^2_S - \gamma_3
\ {\partial \rho_S\over \partial h} -    \gamma_4 \ {\partial
\rho^2_S\over \partial h} \eqno (15) $$ \noindent Here, $
\displaystyle {\gamma_4 \ {\partial \rho^2_S\over\partial h}}$ means
  $ \displaystyle {2 \rho_S \gamma_4 \ {\partial \rho_S\over
\partial h}}  ,$ and
$$ \gamma_1 \ = \ {\mu \pi\over 12 \ \delta^2} \ \rho_2  , \hskip 0,5 cm
\gamma_2 \ = \ { k \pi\over 12 \ \delta^2} \   \eqno (16)$$
$$\displaystyle \gamma_3 \ = \ {\mu \pi\over 6 \ \delta} \ \rho_2
  , \hskip 0,5 cm \gamma_4 \ = \ { k \pi\over 6 } \
\bigg( \ {1\over \sigma} \ - \ {1\over 8 \ \delta} \ \bigg)  $$

\noindent $E_S$ is the form of the special energy to be added at the
solid surface to obtain the total energy of the liquid. In
expression (15), the \ $\gamma_1$ \ term (favoring large \ $\rho_S$)
describes an attraction of the liquid by the solid.
 The \ $\gamma_2$ \ term
represents a  reduction of the liquid/liquid attractive interactions
near the surface : a liquid molecule lying directly on the solid
does not have the same number of neighbors that it would have in
the bulk. The terms with the coefficients  \ $\gamma_3$ \ and \
$\gamma_4$ \  also describe a reduction of the liquid/liquid
attractive interactions due to the lack of molecules of the liquid
near the wall (in the
 case where \ $\displaystyle {\partial \rho_S\over \partial
h}$ \ is positive). \noindent Expressions (15) and (16) do
generalize the results by Nakanishi and Fisher$^{7}$: Expression
(15) contains terms (the first two)  similar to the expression given
by Nakanishi and Fisher, but we find additional terms associated
with \ $\gamma_3$ \ and \ $\gamma_4$. Additional terms are the
correction of the two first terms. Our main interest  is to obtain
values of coefficients of the energy of the
 wall
 as a function of the properties of molecules and to take into account
variations of density at the wall strong enough with respect to
molecular sizes.

\bigskip {\bf 4. Extension to the case of liquid
mixtures in contact with a solid wall}

\bigskip \noindent Here we
propose an extension of the above theory to the case of liquid
mixtures. This example is for a binary mixture, but there is no
reason one could not include more species in the mixture.

The hypotheses are the same as in the case of one-component liquid.
The only difference will come from the interaction of  molecules of
the two liquids.

An adaptation of the previous calculations yields the following
results. The potential energy resulting from the action of all
molecules in the
 medium on molecule $1$ of
 liquid 1 located in $0$ \ is
$$\displaystyle W_{10} \ = \ \int\!\!\!\!\int\!\!\!\!\int_{\Omega_1} \ m^2_1 \ \psi_1
(r) \nu_1 d\omega_1 \  $$
 $$ + \ \int\!\!\!\!\int\!\!\!\!\int_{\Omega_1} \
m_1 m_2 \psi_3 (r)
 \nu_2 d\omega_1 \ +\ \int\!\!\!\!\int\!\!\!\!\int_{\Omega_2} \ m_1 m_3 \chi_1 (r) \nu_3
d\omega_2       \eqno (17)$$ \noindent This energy is only for one
species in the liquid. To determine the whole energy, one must first
sum over all molecules of species $1$ and then do an analogous
summation
 over the molecules of species $2$.
Potential energy \ $W_{10}$ \ will be summed over all  molecules of
the liquid mixture. In this way the contribution of the liquid-liquid
integral, $ \, W_{10}$, will be
 counted twice in the previous
integral over the domain \ $\Omega_1$.

 We have denoted by \ $m_i$,
the mass of the molecule of fluid \ $i   \ ( i \in \lbrace 1 , 2
\rbrace \ ) , \ \psi_1$ \ and \ $\psi_2$ \ are the potentials of
interaction between molecules of fluid $1$ with themselves and among
molecules of fluid $2$, \ $\psi_3$ \  is the potential of molecular
interaction between the two fluids, \ $\chi_i \    ( i \in \lbrace 1
, 2 \rbrace $) are the
 intermolecular
potentials of fluids with the solid wall, and $\nu_i \    ( i \in
\lbrace 1 , 2 , 3 \rbrace $ ) denote
 the number of molecules in
liquids and solid wall per unit volume.

Then, following the procedure of section 2, we obtain :

$$\displaystyle E_{10}\ = \ \nu_{10} W_{10}\ = \ 2 \pi \ \rho^2_1\
\int^{\infty}_{\sigma_1}\ r^2 \psi_1 (r) dr$$ $$ + \ \pi\ \rho^2_1\
\int^{\infty}_h \ r (h-r) \ \psi_1 (r) dr   + \ 2\pi \rho_1 \rho_3
\int^{\infty}_h \ r (r-h) \chi_1 (r) dr $$ $$  + {\pi\over 2}\
\rho_1 \ {\partial \rho_1\over
\partial z}\ \int^{\infty}_h \ r(h^2 - r^2) \ \psi_1 (r) dr
 + \ {\pi\over 3} \ \rho_1 \ \Delta \rho_1 \
\int^{\infty}_{\sigma_1}\ r^4 \psi_1 (r) dr \ $$ $$+ \ {\pi\over
12}\ \rho_1 \ \Delta_T \rho_1 \ \int^{\infty}_h \ (- 2 r^4 + 3 h r^3
- r h^3)\psi_1 (r) dr $$  $$+ \ {\pi\over 6} \ \rho_1 \ {\partial^2
\rho_1\over
\partial z^2}\ \int^{\infty}_h \ r (h^3 - r^3) \psi_1 (r) dr $$  $$+ \
2\ \pi \rho_1 \rho_2\ \int^{\infty}_{\sigma_2}\ r^2 \psi_3 (r) dr
     + \ \pi\rho_1 \rho_2\ \int^{\infty}_h \ r
(h - r ) \psi_3 (r) dr  $$ $$ + \ {\pi\over 2 }\ \rho_1 \ {\partial
\rho_2\over
\partial z}\ \int^{\infty}_h \ r (h^2 -  r^2) \ \psi_3 (r) dr
  + {\pi\over 3} \ \rho_1 \ \Delta \rho_2 \
\int^{\infty}_{\sigma_2}\ r^4 \psi_3 (r) dr  $$
$$ + {\pi\over 12} \
\rho_1 \ \Delta_T \rho_2\ \int^{\infty}_h \ (- 2 r^4 + 3 h r^3 - r
h^3)\psi_3 (r)  dr $$ $$ +  \ {\pi\over 6} \ \rho_1 \ {\partial^2
\rho_2\over
\partial z^2}\ \int^{\infty}_h \ r (h^3 - r^3) \psi_3 (r) dr
 $$

\noindent where \ $\sigma_i   \ ( i \in \lbrace 1 , 2 \rbrace  )$ \
is the diameter of molecule of liquid \ $i$.

We may also repeat the same calculation for \ $E_{20}$ \ associated
with the second component of the mixture. As for a simple fluid, we
may take into account the kinetic effects and the internal energy of
a nonhomogeneous
 mixture as in Fleming et al $^{14}$.
Then, we obtain
 the
following additional energy  at the solid surface for the liquid
mixture in the form
$$\displaystyle E_S \ = \ - \gamma_{11} \ \rho_{1 S} - \gamma_{21} \ \rho_{2 S} +
{1\over 2} \ (\ \gamma_{12} \ \rho^2_{1 S} + \gamma_{22} \ \rho^2_{2
S} + 2 \gamma_{32} \ \rho_{1 S} \rho_{2 S} \ )$$
$$\displaystyle - \gamma_{13} \ {\partial \rho_{1 S}\over \partial h}\
- \gamma_{23}  \ {\partial \rho_{2 S}\over \partial h}\ - \ \bigg(\
\gamma_{14} \ {\partial \rho^2_{1 S}\over \partial h}\ + \
\gamma_{24} {\partial \rho^2_{2 S}\over \partial h}\ + \ 2\
\gamma_{34} \ {\partial \rho_{1 S} \rho_{2S}\over \partial h}\
\bigg)   \eqno(18)$$ \noindent All the coefficients \ $\gamma_{ij}$
\ can be calculated explicitly after the particular form of
interaction potentials was chosen. For example in the case of London
forces, the values of coefficients related to the densities of the
two fluids at the surface are

$$ \gamma_{11} \ = \ {\mu_1 \pi\over 12 \ \delta_1^2} \ \rho_3  , \hskip 0,5 cm
\gamma_{21} \ = \ {\mu_2 \pi\over 12 \ \delta_2^2} \ \rho_3 \ $$
$$\gamma_{12} \ = \ {k_1 \pi\over 12 \ \delta_1^2}  , \hskip 0,5 cm
\gamma_{22} \ = \ {k_2 \pi\over 12 \ \delta_2^2}\  \eqno (19) $$
$$\gamma_{32}\ = \ {k_3 \pi\over {24}}  ( {1\over   \delta_1^2 }+ {1\over  \delta_2^2} )  $$
where $ \rho_3$ is the density of the solid,  $\mu_i  $ are the
coefficients associated with potentials $ \chi_i , \ i \in \lbrace 1
, 2   \rbrace$,  $k_i  $ is associated with potentials $\psi_i, \  i \in \lbrace
1 , 2 , 3  \rbrace $  and \ $\displaystyle\delta_i  = \ {1\over 2}\
(\sigma_i + \tau )$, $ i  \in \lbrace 1 , 2   \rbrace $ are the
minimal distances between the solid and
 molecules of the two species of the mixture.

Such an expression allows one to estimate the influence of a solid
wall on each component of a fluid mixture. Depending on the values
and signs of different coefficients \ $\gamma_{ij}$, one can
estimate the magnitude of the attraction or repulsion effects due to
the wall. Placing the fluid mixture in contact with specially chosen
solid walls may be an efficient way to separate constituents of a
molecular mixture.
\bigskip

{\bf 5. Conclusion }

The energy of a fluid in contact with a solid wall contains a
contribution from the solid which may be represented
 by a surface density function. For a flat wall we obtain this
expression by taking into account not only the density of the liquid
at the surface but also its normal derivative. All the numerical
values of the coefficients of the surface energy functional are
calculated
 in terms of the parameters of molecules in solid and liquid.
 This energy characterizes the behavior of the surface in contact
 with the fluid in the wetting transition.
The method may be extended to the case of nonflat solid surfaces
 which is important in catalysis chemistry.

\bigskip\bigskip
\centerline{{ \bf Appendix 1.}}
\bigskip
\centerline{{\bf Calculation of the value of \ $E$ \ in Eq.
(8)}}\bigskip \noindent To obtain the formula for \ $E$ \
given in Eq. (8), we have to calculate the two integrals:

$$\displaystyle {1\over 2 } \ \int\!\!\!\!\int\!\!\!\!\int_{\Omega_1}
\ m^2_1 \ \psi (r) \nu_1 d\omega_1 \hskip 0,5 cm \hbox{and} \hskip
0,5 cm \int\!\!\!\!\int\!\!\!\!\int_{\Omega_2} \ m_1 m_2 \chi (r)
\nu_2 d\omega_2$$

\noindent a) Calculation of  \ $\displaystyle {1\over 2 } \
\int\!\!\!\!\int\!\!\!\!\int_{\Omega_1} \ m^2_1 \ \psi (r) \nu_1
d\omega_1 $

This integral is associated with the energy of interaction between
molecules in the liquid. \noindent Let us denote by  \ $S (0 , h)$,
  the domain occupied by the sphere centered at
 $(0,0,0)$
\ and with radius \ $h$ \
  (see Figure 1). We introduce \ $(r , \theta , \varphi)$
the spherical coordinates associated with the center of
 the sphere. Then,
 $$ {1\over 2}\ \int\!\!\!\!\int\!\!\!\!\int_{\Omega_1}\ m^2_1\ \psi(r)\
\nu_1\ d\omega_1 = $$ $$ \ {1\over 2}\
\int\!\!\!\!\int\!\!\!\!\int_{S(0,h)}\  m^2_1\ \psi(r)\ \nu_1\
d\omega_1\ + \ {1\over 2}\
\int\!\!\!\!\int\!\!\!\!\int_{\Omega_1-S(0,h)} \ m^2_1\ \psi(r)\
\nu_1\ d\omega_1$$ \noindent Let us note than for any integers $\
p,\ q,\ r,\ $ and any boundary $\ \Sigma\ $ of a sphere
 centered at  $\ (0,0,0)\ $
and radius \ $ \displaystyle r ,$
$$ \int\!\!\!\!\int_{\Sigma}  x^{2p+1}\ y^q\ z^r\ d s =0\quad
 {\rm and} \quad   \int\!\!\!\!\int_{\Sigma}x^2\ d s
 =
\int\!\!\!\!\int_{\Sigma}y^2\ d s  =  \int\!\!\!\!\int_{\Sigma}z^2\ d s
  =   {4\pi\over 3}\ r^4   $$
Then,
$${1\over 2}  \int\!\!\!\!\int\!\!\!\!\int_{S(0,h)} m^2_1  \psi(r)
\nu_1 d\omega_1  =  2  \pi   m^2_1  \nu_{10}  \int^h_{\sigma}
r^2\psi(r)  dr+{\pi\over 3}  m^2_1   \Delta\nu_{10}  \int^h_{\sigma}
r^4 \psi(r)   dr $$ and
$$\int\!\!\!\int\!\!\!\int_{\Omega_1-S(0,h)}  m^2_1\psi(r) \nu_1
d\omega_1  =  {1\over 2}  \int^{2\pi}_{0}\Bigg(
\int^{+\infty}_h\Big( \int^{\pi}_{Arccos  {h\over r}}m^2_1  \psi(r)
\nu_1  \sin   \varphi  d \varphi\Big)   r^2  dr\Bigg)  d\theta$$ The
expansion up to the second order of $\ \nu_1\ $ (see Eq. (4)),
yields
$$ {1\over 2 } \ \int\!\!\!\!\int\!\!\!\!\int_{\Omega_1-S(0,h)}\ m^2_1\
\psi(r)\ \nu_1\ d\omega_1\ =\ \displaystyle\pi\ \nu_{10}\ m^2_1\
\int^{\infty}_h\ r\ (h+r)\ \psi(r)\ dr$$ $$\displaystyle + \ \
{\pi\over 2}\ {\partial\nu_{10}\over \partial z}\ m^2_1\
\int^{\infty}_h \ r\ (h^2-r^2)\ \psi(r)\ dr$$
$$\displaystyle + {\pi\over 12}\ \Bigg( \
{\partial^2\nu_{10}\over \partial x^2} \  + \
{\partial^2\nu_{10}\over \partial y^2}\Bigg)  \ m^2_1\
\int^{\infty}_h \ \Big( 2r^4+3r^3\ h-rh^3\Big)\ \psi (r)\ dr$$
$$ + {\pi\over 6} \ {\partial^2\nu_{10}\over
\partial z^2}\  m^2_1\ \int^{\infty}_h\ r\ (h^3+r^3)\ \psi(r)\ dr$$
\medskip \noindent which is the desired result.
\begin{figure}[h]
\begin{center}
\includegraphics[width=5cm]{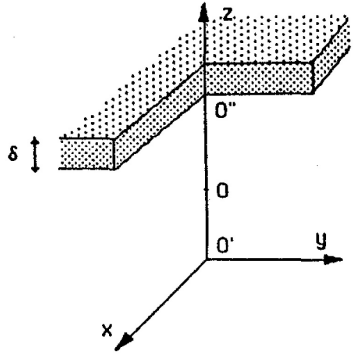}
\end{center}
\caption{\emph{Representation of the changed variable associated with the molecular layer at the solid surface. }}\label{fig2}
\end{figure}

b) Calculation of $\displaystyle \
\int\!\!\!\!\int\!\!\!\!\int_{\Omega_2}\ m_1\ m_2\ \chi (r)\ \nu_2\
d\omega_2$

This term corresponds to the energy with respect to the solid wall.
$$\int\!\!\!\!\int\!\!\!\!\int_{\Omega_2}\ m_1\ m_2\ \chi (r)\ \nu_2\
d\omega_2 \ $$ $$= \ m_1 \ m_2\ \nu_2 \ \int^{2\pi}_0\ \bigg(
\int^{\infty}_h \ \chi (r)\ \ \bigg( \int^{Arccos\ {h\over r}}_0 \
\sin \varphi\ d\varphi \bigg) r^2\ dr \Bigg) \
 d\theta
$$
$$\displaystyle = \ 2\pi\  m_1\, m_2\, \nu_2
\int^{\infty}_h \ r\ (r-h) \ \chi (r)\ dr$$

\noindent which gives relation (7).

 \noindent  Now, we change
variable as in Figure 2.  The origin of the third coordinate $\ z\ $ is
placed at  $ \ 0^{\prime}$. Consequently \ $ z \ = \
\overline{0^{\prime}0}\ $ with $ \ 0 \ $  the position of
 molecule 1.

  If we take now the origin of the $z$-axis at the solid
wall and  change the orientation in such a way that $\
\overline{00^{\prime \prime}}\ =\ h, \ $ we obtain that $\ z+h\ $ is
a constant for  a given molecule.
 Hence, we get
 $$ {\partial\rho_1\over
\partial h}\ = \ - \ {\partial\rho_1\over \partial z}\ ,\ \
{\partial^2\rho_1\over \partial h^2}\ =\ {\partial^2\rho_1\over
\partial z^2}$$
Then, $$ E\ =\ 2\pi\rho^2_1\ \int^{\infty}_{\sigma}\ r^2\ \psi(r)\
dr + \ \pi\rho^2_1\ \int^{\infty}_h\ r (h-r)\ \psi(r)\ dr $$
$$ + 2\pi\rho_1\rho_2\ \int^{\infty}_h\ r (r-h)\ \chi(r)\
dr+{\pi\over 2}\ \rho_1\ {\partial\rho_1\over
\partial h}\ \int^{\infty}_h\ r (h^2-r^2)\ \psi(r) \ dr $$
$$+ {\pi\over 3}\rho_1\Delta\rho_1\ \int^{\infty}_{\sigma}\ r^4\
\psi(r)\ dr+{\pi\over 12}\ \rho_1\Delta_T\rho_1\ \int^{\infty}_h
\Big(-2r^4+3hr^3-rh^3\Big)\ \psi(r) \ dr$$ $$+{\pi\over 6}\rho_1\
{\partial^2\rho_1\over \partial h^2}\ \int^{\infty}_h \ r (h^3-r^3)\
\psi(r)\ dr$$ (we note similarly the Laplace operator in new
coordinates and in old coordinates).
 So, we obtain Eq. (8).

\bigskip
\bigskip
\centerline{ { \bf Appendix 2.}}
 \bigskip
\centerline{{\bf Some remarks about the potential associated with
cohesive forces }}\bigskip \noindent In molecular theory it is
proved by the virial method than the so-called "coefficient a"  of the
van der Waals equation can be obtained from
$$ \int^{\infty}_{\sigma}\ \varphi(r)\ 4\pi r^3\ dr\ =\ -{6a\over
N^2}\eqno{(20)}$$
The term $\ \varphi(r)\ $ is the magnitude of the
attractive force between the two molecules of the fluid at the
distance
 $r,$ and $ N $ is the Avogadro
number. The potential $\displaystyle \ \psi_1(r)\ = \ - \ \int\
\varphi(r)\ dr\ $ is assumed to be zero at infinity.

Take the forces in the form $$ \varphi(r)\ = {A\over r^n}\ ,\ \
\psi_1(r) ={A\over (n-1)r^{n-1}} $$ then,
$$ 4\pi\ A\ \int^{\infty}_{\sigma}\ r^{3-n}\ dr\ =\ {4\pi\ A\ \over n-4}\ {1\over
\sigma^{n-4}}$$  and Eq. $(20) $ \ yields
$$ A \ = \ - \ {3 (n-4) a \ \sigma^{n-4}\over 2 \ \pi \ N^2}\eqno(21) $$
\noindent Let us introduce $\ \psi (r)\ $ such that $\displaystyle\
m^2_1\ \psi(r)\ =\ \psi_1(r)$. The term $\ \psi\ $ gives the
potential per unit mass. Then,
$$\psi(r) \ = \ {-3 (n-4) \ a\ \sigma^{n-4}\over 2(n-1)\ \pi \ N^2 \ m^2_1
\ r^{n-1}} $$ Here, \ $ Nm \ = \ M \ $ denotes the molar mass of the
fluid. In case $\ n\ = 7\ $ (London forces), we get:
$$\psi(r)\ =\  {-3 a\ \sigma^3\over 4 \pi\ M^2\ r^6}$$
In the following, we use:
$$\psi(r)\ =\ - {k\over r^6}\hbox{\hskip 1 cm  with \ \  \ } k\ =\ {3
a\ \sigma^3\over 4 \pi\ M^2}$$ \underbar{Consequence: Calculation of
$ \ E \ $ in Eq. (8) for London forces.}
\bigskip

Now, using direct integration, we obtain the values of the
coefficients given in Eq. (8). Since now
$$\psi (r)\ =\ -\displaystyle{k\over r^{n-1}}\  ,\ \ \ \chi (r)\ =\
-\displaystyle{\mu\over  r^{n-1}}\ ,\ \ \ \ \ \hbox{and} \  \ n\ =\
7$$ \noindent we obtain
$$ \cases{
 \ \ \displaystyle 2\pi\ \int^{\infty}_{\sigma}\ r^2\ \psi(r)\ dr \ = \
- \ { 2 \ k\ \pi\over 3 \ \sigma^3}\cr\cr\cr \ \
\displaystyle{\pi\over 3}\ \int^{\infty}_{\sigma}\ r^4\ \psi(r)\ dr\
= \ -\ {k\ \pi\over 3\ \sigma}\cr\cr\cr \ \ \displaystyle 2\ \pi \
\int^{\infty}_{h}\ r\ (r-h)\ \chi(r)\ dr\ =\ - \ {\mu\ \pi\over 6\
h^3}\cr\cr\cr \ \ \displaystyle \pi\ \int^{\infty}_{h}\ r\ (h-r)\
\psi(r)\ dr\ = \  {k\ \pi\over 12 \ h^3}\cr\cr \cr \ \ \displaystyle
-\ {\pi\over 2} \int^{\infty}_{h}\ r\ (h^2-r^2)\ \psi(r)\ dr\ =\
-\displaystyle{k\ \pi\over 8 \ h^2}\cr\cr \cr \ \
\displaystyle{\pi\over 12} \int^{\infty}_{h}\ (-2r^4+3h\ r^3-r\
h^3)\ \psi(r)\ dr \ = \ {k \ \pi\over 16 \ h}\cr\cr \cr \ \
\displaystyle {\pi\over 6} \int^{\infty}_{h}\ \ r\ (h^3-r^3)\
\psi(r)\ dr\ =\ \ \ \displaystyle{k \ \pi\over 8\ h} }$$ \vskip 1.5
cm {\bf  Acknowledgements:} I have greatly benefited from generous
discussions and correspondence with Professor
 B. Widom
and D$^{\hbox{r}}$ A. E. van Giessen. Without their interest and
help, this paper would be never published. The support of PRC/GdR
CNES-CNRS 1185 is gratefully acknowledged. \vskip 3.5  cm
\underbar{\bf References}

\medskip
 \noindent (1) Cahn   J. W., J. Chem. Phys.,
{\bf 1977},  66,  3667.

 \noindent  (2) van Giessen   A. E., Bukman  D. J., Widom  B.,  J.  Colloid and Interface Science, {\bf 1997}, 192,  257.

 \noindent  (3)  Zisman   W. A., In "Contact angle, wettability and adhesion":
 Advances in Chemistry Series,  43 (Gould R. F., ed.), A.C.S., Washington D. C.,
{\bf 1964}, 1.

 \noindent (4) Li  D., Neuman  A. W., Langmuir {\bf 1993}, 9, 3728.

 \noindent (5)   Kwok  D. Y., Li  D., Neuman  A. W., Colloids surfaces, {\bf
1994}, 89,  181.

 \noindent  (6)  Rowlinson  J. S.,  Widom  B.,
Molecular theory of capillarity, Clarendon Press, Oxford, {\bf
1984}.

 \noindent (7) Nakanishi  H.,  Fisher  M.E., Phys. Rev. Lett., {\bf 1982}, 49,
1565.

 \noindent (8) de Gennes  P. G., Review of Modern Physics, {\bf 1985}, 57, 3,
  827.

 \noindent (9) Snook I.,  van Megen  W., J.  Chem. Phys., {\bf 1979}, 70, 3099.

 \noindent (10) Israelachvili   J., Intermolecular and surface forces, Academic
Press, London, {\bf 1992}.

 \noindent (11) Rocard  Y., Thermodynamique, Masson, Paris, {\bf 1967}, Chapter
5.

 \noindent
(12) Gouin  H., Comptes Rendus Acad. Sci. Paris, {\bf 1988}, 306, II,
755.

 \noindent
(13) de Gennes  P. G., J. Phys. (Paris) Lett. {\bf 1981}, 42, 377.

 \noindent
(14) Fleming   P. D.,  Yang  A. J. M.,  Gibbs  J. H. J., Chem. Phys.,
{\bf 1976}, 65, 7.

\end{document}